\PassOptionsToPackage{sort}{natbib}

\documentclass[Afour,sageh,times,doublespace]{sagej}

\usepackage{moreverb,url}

\usepackage[colorlinks,bookmarksopen,bookmarksnumbered,citecolor=red,urlcolor=red]{hyperref}

\usepackage{booktabs}
\usepackage{tabularx}
\usepackage{array}

\newcommand\BibTeX{{\rmfamily B\kern-.05em \textsc{i\kern-.025em b}\kern-.08em
T\kern-.1667em\lower.7ex\hbox{E}\kern-.125emX}}

\makeatletter

\def\@maketitle{%
  \vspace*{-30pt}%
  \null%
  \begin{center}
  \begin{minipage}[t]{\textwidth}
    \vskip 12pt%

    {\raggedright
      \titlesize
      \textbf{\@title}
      \par
    }

    \vskip 12pt%

    {\raggedright
      \normalsize
      \textbf{\@author}
      \par
    }

    \vskip 24pt%

    {\noindent\usebox\absbox\par}

    \vspace{18pt}%

    {\noindent\normalsize\@keywords\par}

  \end{minipage}
  \end{center}

  \vspace{18pt}%
  \par
}

\def\ps@title{%
  \def\@oddhead{%
    \parbox{\textwidth}{%
      \mbox{}\\[-1pt]%
      \noindent\rule{\textwidth}{0.5pt}%
    }%
  }%
  \let\@evenhead\@oddhead
  \def\@oddfoot{}%
  \let\@evenfoot\@oddfoot
}

\def\ps@sagepage{%
  \let\@mkboth\@gobbletwo

  \def\@evenhead{%
    \parbox{\textwidth}{%
      \normalsize\sagesf
      \thepage\hfill
      \\[-6pt]
      \noindent\rule{\textwidth}{0.25pt}%
    }%
  }%

  \def\@oddhead{%
    \parbox{\textwidth}{%
      \normalsize\sagesf
      \hfill\thepage
      \\[-6pt]
      \noindent\rule{\textwidth}{0.25pt}%
    }%
  }%

  \def\@evenfoot{}%
  \let\@oddfoot\@evenfoot
}

\makeatother

\begin{document}

% \runninghead{}

\title{Causal Humility: Governing the Authority of Fairness Claims Under Causal Uncertainty}

\author{%
Khadija Zanna\affilnum{1},
Rodrigo Ferreira\affilnum{1},
Frederick L. Oswald\affilnum{2},
and Akane Sano\affilnum{1}%
}

\affiliation{%
\affilnum{1}Rice University, Houston, Texas, USA\\
\affilnum{2}University of California, Irvine, Irvine, California, USA%
}

\email{khzanna@rice.edu}

\begin{abstract}

AI systems increasingly shape consequential decisions in employment, lending, healthcare, and public services, making fairness an important problem for auditing and governance. Fairness assessment often requires more than identifying disparities in outcomes; it also requires understanding how those disparities arise. Causal approaches to algorithmic fairness address this explanatory problem by representing pathways through which protected attributes and other factors may influence outcomes. Yet causal reasoning in sociotechnical systems remains uncertain: relevant confounders may be unobserved, causal directions contested, and conclusions sensitive to modeling assumptions and representational choices.

We introduce \emph{causal humility} as a governance framework for managing the relationship between epistemic uncertainty and \emph{causal authority}. Drawing on causal fairness research, epistemology, political philosophy, and science and technology studies, \emph{causal humility} has four requirements: making consequential assumptions and omissions visible; preserving contestability across defensible causal accounts; recognizing indeterminacy as a legitimate analytical outcome; and maintaining proportionality between evidentiary support and institutional authority or action.

We operationalize these requirements through \emph{Causal Cards}, a documentation artifact for recording the causal basis, assumptions, omissions, model dependence, and governance implications of fairness claims. Using the UCI Adult dataset as a worked epistemic demonstration, we show how different causal discovery procedures applied to the same observational data render fairness-relevant pathways visible, absent, or unresolved. \emph{Causal humility} treats such disagreement as information about the limits of what can responsibly be claimed and the authority those claims should carry.

\end{abstract}


\maketitle

\pagestyle{sagepage}

\section{Introduction}
\label{sec:introduction}

AI systems increasingly shape access to employment, credit, healthcare, education, and public services \citep{panda2025artificial}. As their use becomes more institutionally consequential, organizations face growing pressure to justify whether automated decisions are fair, explainable, and ethically or legally defensible \citep{calota2025impact}. Evaluating fairness, however, often requires knowing not only \emph{that} outcomes differ, but \emph{why} they differ \citep{kusner2017counterfactual, kilbertus2017avoiding, sen2009idea}. Fairness assessments therefore produce explanations of whether disparities reflect discrimination, legitimate qualification, historical disadvantage, institutional structure, measurement choices, or some combination of these mechanisms.

Causal approaches to algorithmic fairness make this explanatory ambition explicit. Rather than characterizing disparities solely through observed inputs and outcomes, causal approaches seek to distinguish among the mechanisms through which those disparities arise, including whether protected attributes influence outcomes directly or through pathways regarded as legitimate or impermissible \citep{carey2022causal}. Causal reasoning thus offers an account of why a disparity exists and which processes should matter to its evaluation. This is especially important given longstanding critiques of fairness approaches that abstract technical assessments from the social and institutional processes producing inequality \citep{green2018myth, selbst2019fairness}. We approach causal fairness as a \emph{data practice} through which observational data, modeling assumptions, variable representations, and normative classifications are assembled into explanations of disparity.

Yet causal explanation also carries greater interpretive consequence. Once a causal account is used to evaluate fairness, it can shape which explanations of inequality become visible, which pathways are treated as discriminatory or legitimate, where responsibility is located, and what interventions appear warranted \citep{zanna2026uncovering}. A causal model can therefore become more than a technical representation of relationships among variables. When its outputs inform an audit or governance process, they can contribute to institutional claims about what is fair, unfair, remediable, or acceptable.

We refer to this as \emph{causal authority}: the epistemic and institutional force granted to a causal explanation when it is used to classify disparities, allocate responsibility, justify intervention, or support judgments about an AI system. \emph{Causal authority} is not inherently problematic. The concern arises when the authority attached to a causal fairness claim exceeds the epistemic support of the causal account on which it depends.

This possibility is especially important in sociotechnical settings, where causal knowledge is often underdetermined. Relevant confounders may be unavailable or unobserved; causal directions may be ambiguous or contested; variables may plausibly occupy different causal roles; and causal discovery (CD) procedures can recover divergent structures from the same observational evidence \citep{glymour2019review, cartwright2007hunting}. Choices about variable inclusion, proxy construction, pathway permissibility, and the treatment of unobserved social processes can, therefore, materially shape the resulting causal account. Recent work on graph uncertainty and imperfect causal structures further demonstrates that causal fairness conclusions can depend substantially on assumptions about the underlying model \citep{valerio2026counterfactual, su2026counterfactual}.

These conditions create a governance problem as well as a technical one. If a fairness claim remains dependent on partial, contested, or assumption-sensitive causal evidence, what authority should that claim be permitted to carry? When is the evidence strong enough to support a categorical conclusion? And when should disagreement or indeterminacy remain part of the institutional judgment? These questions matter as causal approaches are developed for fairness auditing and evaluation, including settings involving uncertain or incompletely observed causal structure \citep{fang2024peer, nagesh2026evaluating, byun2024auditing}.

We introduce \emph{causal humility} as a governance principle for addressing this problem. \emph{Causal humility} does not reject causal analysis, treat all causal models as equally credible, or require action to be postponed until uncertainty disappears. Rather, it holds that the strength of a fairness claim and the authority granted to it should remain proportionate to the stability and evidentiary support of the causal reasoning on which it depends. We develop this principle around four requirements: making consequential assumptions and omissions visible; preserving contestability across defensible causal accounts; recognizing indeterminacy as a legitimate analytical outcome; and maintaining proportionality between epistemic support and institutional action.

We operationalize these requirements through \emph{Causal Cards}, a structured documentation artifact that records the causal representation underlying a fairness claim, consequential assumptions and omissions, the stability of relevant findings across alternative specifications, and the governance response associated with their epistemic status. This serves as a method to document uncertainty and reduce the likelihood that material qualifications disappear as causal findings move from technical analysis into institutional decision-making.

We illustrate the framework using the UCI Adult dataset as a worked epistemic demonstration. Applying three CD procedures to the same observational feature set, we show how alternative causal representations can render fairness-relevant pathways visible, absent, or unresolved. The example is not intended to establish a definitive causal account of income inequality or validate a production audit. Instead, it demonstrates how causal representation can structure what analysts are in a position to claim before a categorical fairness judgment is made.

Our contributions are threefold:
\begin{enumerate}
    \item We introduce \emph{causal authority} to characterize the epistemic and institutional force acquired by causal explanations when they support consequential fairness claims, and identify mismatches between \emph{causal authority} and epistemic support as a governance problem.

    \item We develop \emph{causal humility} as a governance principle for maintaining an accountable relationship between uncertain causal evidence and the authority granted to resulting fairness claims.

    \item We operationalize \emph{causal humility} through \emph{Causal Cards} and a worked case showing how disagreement among causal representations can alter which fairness-relevant explanations become visible, absent, or unresolved.
\end{enumerate}

The remainder of the paper examines causal fairness as a knowledge practice, develops \emph{causal authority} under epistemic underdetermination, introduces \emph{causal humility} and \emph{Causal Cards}, and uses the Adult Income example to examine the implications of model disagreement for fairness interpretation and governance.

\section{Causal Fairness as a Knowledge Practice}
\label{sec:causal_fairness_knowledge}

To understand why uncertainty in causal fairness becomes a governance problem, it is first necessary to consider what causal fairness analysis is intended to produce. As a data practice, fairness analysis produces categories and explanations through which observed disparities become interpretable. However, statistical and causal approaches differ in the kinds of claims they support. Statistical criteria characterize patterns in observed outcomes, whereas causal approaches seek to identify the mechanisms through which those patterns arise. This move from description to explanation expands what fairness analysis can say, but it also increases the epistemic demands placed on the models through which those explanations are produced.

\subsection{Statistical Fairness and the Turn to Causal Explanation}

Early work on fairness in ML largely operationalized fairness through statistical criteria defined over model inputs and outputs. Measures such as demographic parity, equalized odds, equal opportunity, and calibration provide formal ways to characterize whether groups receive systematically different treatment or outcomes \citep{chouldechova2017fair, pessach2022review, verma2018fairness}. These approaches have generated an extensive literature on the mathematical compatibility, trade-offs, and limitations of different fairness definitions \citep{chouldechova2017fair}.

Statistical criteria are valuable precisely because they make disparities measurable without requiring a complete account of how these disparities arise. At the same time, this feature limits the kinds of fairness claims they can support. Observing that two groups experience different outcomes does not by itself establish whether the difference reflects discrimination, prior disadvantage, measurement error, unequal opportunity, legitimate qualification, or some combination of these mechanisms. Moreover, the criteria of statistical fairness themselves encode normative choices about which distributions, errors, or forms of treatment should be equalized \citep{carey2023statistical}. Critical scholarship has therefore cautioned against treating fairness as a property that can be evaluated independently of the social and institutional processes that generate the observed data \citep{green2018myth,selbst2019fairness}.

Causal approaches respond to this explanatory limitation by asking how disparities are produced. Structural causal models, counterfactual fairness, path-specific effects, and related methods seek to distinguish direct from mediated effects and, in some formulations, permissible from impermissible pathways through which protected attributes may influence outcomes \citep{kilbertus2017avoiding,kusner2017counterfactual,makhlouf2020survey,makhlouf2024causality}. Therefore, causal fairness can go beyond asking whether an outcome differs across groups and can also ask whether the disparity would persist under an intervention, whether a protected attribute influences an outcome through a particular mediator, or whether removing a specified causal pathway would alter the resulting decision.

This shift is consequential because explanation changes the object of fairness evaluation. A statistical disparity is evidence of a difference that requires interpretation. A causal account offers an explanation of the processes through which that difference emerged. Such explanations can support claims about which mechanisms are relevant to discrimination, which variables should be treated as legitimate mediators, and where intervention should occur. Causal reasoning thus expands the analytical vocabulary of fairness, but it also introduces a stronger dependence on the validity of the causal representation through which those claims are generated.

\subsection{The Epistemic Instability of Causal Fairness}

The additional explanatory power of causal fairness comes with substantial epistemic requirements. Causal claims generally depend on assumptions about causal structure, identification, measurement, and the absence or treatment of confounding \citep{pearl2009causality,bareinboim2016causal}. Yet these conditions are difficult to establish in many sociotechnical settings, where relevant social processes are only partially observed and available data are typically observational rather than experimental.

One source of instability is unobserved confounding. Fairness-relevant relationships may depend on mechanisms that are absent from the available data, such as prior discrimination, institutional exclusion, differential access to resources, or other social processes that jointly influence observed variables and outcomes. \citet{fawkes2024fragility}, for example, show that causal fairness conclusions can be sensitive to assumptions about unobserved confounding, while related work develops sensitivity-based approaches for evaluating fairness when causal relationships are incompletely identified \citep{ICLR2024_b795fcfe}. More recent work similarly examines counterfactual fairness when the underlying causal graph is imperfect or uncertain \citep{valerio2026counterfactual, su2026counterfactual}. Collectively, these approaches demonstrate that a fairness conclusion may depend not only on observed disparities but on assumptions about mechanisms that cannot be directly recovered from the available data.

A second source of instability concerns causal structure itself. Many causal fairness formulations presuppose that a relevant causal model has been specified or can be estimated with sufficient confidence \citep{zhang2018fairness, makhlouf2024causality}. However, when that structure is learned from observational data, different CD methods rely on different assumptions regarding conditional independence, latent confounding, acyclicity, functional form, and model selection \citep{glymour2019review,heinze2018causal,malinsky2018causal}. As a result, different procedures can recover different graphs from the same data, while multiple structures may remain observationally compatible with the evidence.

This problem is not unique to CD. More broadly, underspecification occurs when multiple models are compatible with available observations while supporting different interpretations or downstream decisions \citep{d2022underspecification}. Within causal fairness, the consequences are particularly important because differences in structure can alter which pathways are identified as direct or mediated, which variables appear to transmit the effects of protected attributes, and whether particular fairness criteria can be evaluated at all. Recent empirical work illustrates this dependence by comparing CD procedures for path-specific fairness and showing that structural choices can affect downstream fairness analysis \citep{nagesh2026evaluating}.

A third source of instability concerns the assignment of meaning to variables and pathways. Causal models require analysts to decide which constructs to measure, which variables to include, and how those variables should be positioned in relation to one another. A variable such as education, occupation, employment history, healthcare utilization, or neighborhood can be interpreted differently depending on the social context and fairness question under consideration. The same variable might be treated as a legitimate mediator in one causal account and as a proxy for accumulated disadvantage in another. Similarly, determining whether a pathway should count as permissible is not solely a question of statistical identification; it can require a normative judgment about which mechanisms of influence are socially or ethically acceptable \citep{carey2022causal,makhlouf2024causality}.

These forms of instability should not be confused. Uncertainty about an unobserved confounder differs from uncertainty about edge orientation, and both differ from normative disagreement over whether an identified pathway should be considered legitimate. Some uncertainties may be reduced through better data, stronger identification strategies, domain expertise, or sensitivity analysis. Others arise because available evidence does not uniquely settle the representation of complex social processes or the normative interpretation attached to them. What they share is that the resulting fairness claim is conditional on a set of choices and assumptions whose status may be more uncertain than the final claim suggests.

\subsection{From Causal Models to Institutional Claims}

The significance of this instability depends on what causal models are expected to do. If a causal graph is used only as an exploratory research hypothesis, disagreement among alternative structures may remain an ordinary methodological concern. The stakes change when causal analysis is used to support a fairness assessment. At that point, the model can contribute to claims about why a disparity exists, which pathways should be scrutinized, whether an observed difference is normatively concerning, and what forms of remediation appear appropriate.

This transition from model to claim is important because causal representations are selective. A graph foregrounds particular variables and relationships while leaving others unrepresented. An edge can make one mechanism available as an explanation; its absence can make that same mechanism analytically invisible. A mediator classified as legitimate can transform the interpretation of a protected attribute's relationship with an outcome, while an unresolved edge can prevent a corresponding causal claim from being identified. The causal model therefore structures the space of explanations that a fairness analysis can produce.

Existing technical work recognizes the need to state causal assumptions and examine robustness across specifications \citep{pearl2009causality, fawkes2024fragility, ICLR2024_b795fcfe}. Our concern begins one step later: with what happens when conclusions generated under those assumptions are translated into claims that can circulate beyond the technical analysis itself. For example, a finding that a protected attribute has no identified causal pathway to an outcome may be technically conditional on a particular graph, set of measured variables, and identification assumptions. However, once summarized as evidence that the protected attribute ``does not causally affect'' the outcome, the conditional nature of that result can become less visible.

Causal fairness therefore operates as a data practice through which data selection, modeling assumptions, causal representations, and normative classifications are assembled into knowledge claims about how disparities arise and what those disparities signify. The knowledge produced by this process is consequential because it can inform how subsequent actors understand responsibility, legitimacy, and the need for intervention. This does not imply that causal explanations are arbitrary or that causal inference cannot produce reliable knowledge. It means that the epistemic status of the resulting claim depends on the assumptions, evidence, and representational choices through which it was produced.

This distinction motivates the next step in our argument. Technical uncertainty becomes a governance problem when the institutional force granted to a causal fairness claim is disconnected from the conditional and situated basis of the knowledge that produced it. We therefore turn next to \emph{causal authority under epistemic underdetermination}, looking at how causal representations make some explanations of inequality available while excluding others, how omissions and representational choices intersect with structural inequality, and why the authority accorded to causal claims requires governance when their foundations remain partial or contested.

\section{\emph{Causal Authority} Under Epistemic Underdetermination}
\label{sec:causal_authority}

The preceding section characterized causal fairness as a knowledge practice: causal models organize explanations of how disparities arise and what those disparities signify. This explanatory role becomes consequential when causal claims are used to distinguish legitimate from discriminatory mechanisms, assign responsibility, or justify intervention. Yet the authority of these claims may exceed what the available evidence can uniquely establish. We describe this as a problem of \emph{causal authority under epistemic underdetermination}: causal representations can acquire institutional force even when the social processes they represent remain only partially observed, contestable, or compatible with multiple plausible accounts.

Epistemic underdetermination is broader than ordinary statistical uncertainty. It concerns situations in which available evidence does not uniquely determine the causal representation through which a fairness problem is interpreted. Relevant mechanisms may be unobserved, social constructs may admit multiple operationalizations, and normative judgments may be required to determine which pathways matter for fairness. A causal model can remain useful under these conditions, but the authority granted to its conclusions should reflect these representational limits.

\subsection{Situated Causal Knowledge}

STS and social epistemology provide a useful lens for understanding why causal representations should not be treated as neutral mirrors of social reality. Haraway's account of situated knowledge emphasizes that knowledge is produced from particular positions and systems of observation rather than from a disembodied ``view from nowhere'' \citep{haraway2013situated}. Harding similarly draws attention to how scientific inquiry is shaped by decisions about which questions are asked, which evidence is considered relevant, and whose perspectives are incorporated \citep{harding1991whose}.

Applied to causal fairness, situatedness does not imply that causal claims are merely subjective. Empirical evidence, substantive theory, temporal ordering, and domain knowledge can constrain competing causal accounts. Rather, situatedness directs attention to the conditions under which those accounts are produced. Analysts must decide what outcome to model, which variables and constructs to include, how they are operationalized, and what forms of knowledge inform the structure. These choices shape which causal explanations subsequently become available.

However, once encoded in a causal graph, such choices can become less visible. Latour's account of ``black boxing'' describes how judgments and assumptions embedded in technical objects can recede from view once those objects become stabilized and routinely used \citep{latour1987science}. Related work on ML ground truth similarly shows how apparently objective representations are produced through prior decisions about measurement, categorization, and institutional practice \citep{jaton2021constitution, kang2023ground}. A causal graph can undergo a similar transformation: representational choices may come to appear as properties of the world rather than features of a particular modeling process.

This also raises the question of whose knowledge informs causal specification. Affected communities and domain experts may possess knowledge about discrimination, institutional processes, or informal mechanisms that is poorly represented in administrative datasets. Intersectional and participatory work in ML has shown how choices about data, labels, and problem formulation can privilege some forms of knowledge while rendering others marginal \citep{d2023data,klein2024data, klumbyte2022critical, suresh2022towards}. In causal fairness, these choices matter not only for prediction but for which explanations of inequality can be formally represented.

\subsection{Omission, Representation, and Structural Inequality}

These representational limits are especially important when fairness questions concern structural inequality. Many mechanisms relevant to discrimination and disadvantage, including historical exclusion, labor-market segmentation, institutional discrimination, inherited wealth, unequal access to education or healthcare, and neighborhood segregation may shape observed outcomes without appearing explicitly in available datasets \citep{selbst2019fairness, marmot2004status, obermeyer2019dissecting}.

All causal models necessarily simplify. The concern arises when an omitted mechanism is consequential to the fairness interpretation while its absence from the model is treated as evidence of irrelevance. For example, if a dataset contains educational attainment but not unequal access to educational opportunity, education may appear as an individual mediator while the institutional processes that produced educational differences remain outside the causal account. Likewise, employment history may appear as evidence of qualification while prior discriminatory hiring practices or unequal caregiving burdens remain unobserved.

Absence from a causal representation is therefore not equivalent to absence from the social process being modeled. Nor do proxies straightforwardly solve this problem. Variables such as neighborhood, occupation, healthcare utilization, or educational attainment may reflect broader structural processes without being equivalent to them, and different operationalizations can support different causal interpretations.

These choices affect what is permitted to count as an explanation. Critical work on abstraction and intersectionality has shown how technical representations can obscure social context and privilege dominant forms of classification \citep{selbst2019fairness, d2023data, klein2024data}. Fricker's account of epistemic injustice further highlights how exclusion can affect whose explanations are recognized as credible \citep{fricker2007epistemic}. Causal representations inevitably leave some social mechanisms unmodeled. When consequential mechanisms remain outside the representation, their omission should constrain the authority of claims that depend on their assumed irrelevance.

A causal model may therefore be internally coherent and technically well-supported while remaining a partial representation of the social processes relevant to the fairness question. Technical adequacy within the model does not by itself establish the completeness of the explanation derived from it.

\subsection{Why Uncertainty Becomes a Governance Problem}

Epistemic uncertainty becomes a governance problem when causal findings are used to support institutional judgments. An uncertain graph used for exploratory research carries different consequences from the same graph used to justify a conclusion that a disparity is legitimate, mitigation is unnecessary, or a system has satisfied a fairness review. The relevant issue is, therefore, how much authority is granted to the claim produced under uncertainty.

Technical responses such as sensitivity analysis, robustness checks, or additional data can help characterize or reduce uncertainty \citep{fawkes2024fragility, ICLR2024_b795fcfe}. However, they do not determine what an institution is entitled to claim or what action should follow. A finding may remain conditional on assumptions even after careful analysis, and the institutional question then becomes whether those conditions are material to the conclusion being asserted.

This creates a problem of proportionality between \emph{epistemic support} and \emph{causal authority}. A claim that remains stable across well-supported alternatives has a different evidentiary basis from one that depends on a contested modeling choice or unresolved causal direction. The governance concern arises when these differences disappear as causal findings are converted into categorical fairness judgments.

Recognizing this mismatch does not imply that all causal models are equally credible or that institutions should refrain from acting under uncertainty. Evidence, theory, domain expertise, and sensitivity analysis can rule out weak alternatives, and consequential decisions often must be made without complete certainty. The point is that residual uncertainty that is material to a fairness claim should remain consequential to the authority granted to that claim.

This motivates \emph{causal humility}: a governance principle for maintaining an accountable relationship between the evidentiary support of a causal explanation and the institutional force of the fairness claim derived from it.

\section{\emph{Causal Humility} as a Governance Principle}
\label{sec:causal_humility}

The preceding sections identified a potential mismatch between the epistemic support available for a causal explanation and the institutional authority granted to it. We use the term \emph{causal humility} to describe a governance principle for managing this relationship. It requires that the strength and scope of a causal fairness claim remain responsive to what the underlying analysis establishes, assumes, and leaves unresolved.

\emph{Causal humility} does not imply skepticism toward causal analysis or require equal consideration of every conceivable model. Empirical evidence, domain knowledge, temporal constraints, and substantive theory can rule out weak alternatives. Its concern is residual uncertainty that remains material to a consequential fairness claim. We develop the principle around four requirements: assumption disclosure, contestability, recognition of indeterminacy, and proportionality between evidence and action.

\subsection{Assumption Disclosure}

The first requirement is to make consequential causal assumptions, omissions, and modeling choices visible. These may concern causal structure, variable inclusion, unobserved confounding, measurement, variable roles, or the classification of pathways as permissible or impermissible \citep{carey2022causal,makhlouf2024causality}. The relevant assumptions are those whose alteration could materially change the interpretation of a disparity or the claim made from it.

This extends familiar practices of technical documentation \citep{mitchell2019model,gebru2021datasheets} to the explanatory basis of fairness claims. If a conclusion depends on a disputed treatment of a mediator or on the omission of a potentially important social mechanism, that dependency should remain visible in how the finding is stated and used.

\subsection{Contestability and Model Disagreement}

The second requirement is that causal fairness claims remain open to challenge when defensible alternatives support materially different interpretations. The relevant comparison set need not include every possible graph or assumption; it should be constrained by empirical evidence, substantive knowledge, and the purpose of the analysis. However, where multiple plausible alternatives remain, the stability of the fairness-relevant conclusion across them becomes part of the evidence.

Contestability may concern different causal graphs, assumptions about unobserved confounding, competing operationalizations of social constructs, or disagreement over pathway permissibility. Sensitivity analysis can help identify which assumptions drive changes in the result \citep{fawkes2024fragility, ICLR2024_b795fcfe}. Under \emph{causal humility}, such disagreement also determines how much authority the resulting fairness claim can reasonably carry. Domain experts, auditors, and affected stakeholders may also identify omitted mechanisms or contested interpretations that are poorly represented in the observed data. Such challenges need not determine the causal model, but they can expose assumptions relevant to the authority of the resulting claim.

\subsection{Indeterminacy as a Legitimate Outcome}

The third requirement is recognition of \emph{indeterminacy} as a legitimate analytical outcome. Fairness assessments often create pressure for categorical conclusions, yet the available evidence may not always support a determinate causal judgment. An analysis may identify a disparity and several plausible mechanisms while remaining unable to establish which explanation sufficiently accounts for it.

Identifying the limits of what the evidence supports can itself be an analytically informative result. ``Unresolved under the available causal evidence'' can be more accurate than a categorical conclusion produced by suppressing material uncertainty.

An indeterminate finding should specify what remains unresolved, why that uncertainty matters to the fairness claim, and what additional evidence or analysis could reduce it where possible. This makes indeterminacy accountable and contestable.

\subsection{Proportionality Between Evidence and Action}

The fourth requirement concerns the relationship between epistemic support and institutional authority. A causal conclusion that remains stable across well-supported alternatives can reasonably support a stronger claim than one that changes across defensible specifications or depends on unresolved structure.

The appropriate response also depends on the stakes of the decision, the severity and reversibility of potential harms, the availability of alternatives, and the capacity for monitoring and revision. In high-stakes settings, uncertainty about serious discriminatory harm may justify additional precaution rather than inaction.

\emph{Causal humility} instead requires decision-makers to make explicit how the epistemic status of the evidence affected the authority granted to the resulting claim and the action taken on its basis. Its central demand is accountability for the relationship between evidence, claim, and institutional action. The next section operationalizes this relationship through \emph{Causal Cards}.

\section{Operationalizing \emph{Causal Humility}: \emph{Causal Cards}}
\label{sec:causal_cards}

The principles of \emph{causal humility} require a mechanism for carrying the epistemic conditions of a causal fairness claim beyond the technical analysis in which it was produced. Assumptions, omissions, sensitivity results, and unresolved disagreements may be visible to analysts but become less apparent as findings are summarized for auditors, compliance teams, or organizational decision-makers. We introduce \emph{Causal Cards} as a structured documentation artifact intended to preserve this information as causal claims move across technical and institutional contexts.

\emph{Causal Cards} are not intended to determine whether a causal model is correct or to replace sensitivity analysis, robustness checks, domain expertise, or substantive fairness evaluation. Their purpose is to make explicit the relationship among three elements that are often separated in practice: the \emph{causal representation} on which a fairness claim depends, the \emph{epistemic status} of the resulting finding, and the \emph{institutional action} justified on its basis.

\subsection{Positioning \emph{Causal Cards} in a Governance Workflow}

A \emph{Causal Card} is intended to accompany a causal fairness assessment when technical findings are translated into an audit, review, or governance judgment. A stylized workflow is:

\begin{enumerate}
    \item causal model or model set
    \item fairness-relevant analysis
    \item Causal Card
    \item institutional review and action
\end{enumerate}

Analysts are best positioned to document the assumptions, specifications, and limitations underlying the causal analysis. However, judgments about pathway permissibility, acceptable uncertainty, or appropriate institutional response may also require input from domain experts, auditors, legal or policy teams, and affected stakeholders. The final governance determination should therefore remain distinct from the technical analysis rather than being implicitly delegated to the causal model.

This separation is central to \emph{causal humility}. A technical analysis may show that a finding is stable under a defined set of assumptions without determining whether the remaining risk is institutionally acceptable. Conversely, an institution may decide to proceed despite residual uncertainty, without making the underlying causal claim more certain. The \emph{Causal Card} is intended to preserve this distinction.

\subsection{Core Components of a Causal Card}

A \emph{Causal Card} records six categories of information: the decision context and fairness claim; the causal representation considered; consequential assumptions and omissions; the stability and contestability of the finding; the epistemic status of the resulting claim; and the governance response associated with that status. Table~\ref{tab:causal_card_structure} summarizes these components.

\begin{table*}[ht]
\centering
\small
\caption{Core components of a \emph{Causal Card}.}
\label{tab:causal_card_structure}
\begin{tabularx}{\textwidth}{p{0.23\textwidth} X}
\toprule
\textbf{Component} & \textbf{Information recorded} \\
\midrule

Decision context and fairness claim &
The decision setting, outcome of interest, fairness-relevant attributes, population, intended use of the analysis, and the specific causal fairness claim being evaluated. \\

\addlinespace

Causal representation &
The causal model or model set considered, including whether the structure is expert-specified, learned from data, partially identified, or constructed through multiple methods. Relevant substantive constraints on the model set should also be stated. \\

\addlinespace

Consequential assumptions and omissions &
Assumptions about causal structure, variable roles, confounding, measurement, directionality, and pathway permissibility that materially affect the fairness interpretation. Known omissions, suspected unobserved mechanisms, and relevant proxies should also be recorded. \\

\addlinespace

Stability and contestability &
Whether the fairness-relevant finding persists across plausible alternative models, sensitivity assumptions, variable interpretations, or pathway classifications, and which choices drive material disagreement. \\

\addlinespace

Epistemic status &
A characterization of the resulting claim as comparatively \emph{stable}, \emph{assumption-sensitive}, or \emph{unresolved} under the alternatives considered. These categories describe evidentiary status rather than universal numerical thresholds. \\

\addlinespace

Governance response &
The action taken or recommended in light of the finding, its uncertainty, and the stakes of the decision, including where relevant additional review, mitigation, conditional authorization, monitoring, restricted use, or refusal to issue a categorical fairness determination. \\

\bottomrule
\end{tabularx}
\end{table*}

Two design choices are particularly important. First, epistemic status attaches to a \emph{specific fairness claim}, not to a causal graph as a whole. The same representation may provide relatively strong support for one relationship while leaving another unresolved. Second, the categories \emph{stable}, \emph{assumption-sensitive}, and \emph{unresolved} are descriptive rather than fixed decision thresholds. Their purpose is to make explicit how strongly the evidence supports the claim at issue and which assumptions materially affect it.

\subsection{From Documentation to Governance}

\emph{Causal Cards} build on documentation practices such as Model Cards and Datasheets, which record model characteristics, intended uses, data provenance, and known limitations \citep{mitchell2019model, gebru2021datasheets}. Their object of documentation is different: a \emph{Causal Card} records the \emph{explanatory structure through which a fairness claim is justified}. It therefore focuses on which causal account made the claim possible, what remains outside that account, and how dependent the conclusion is on unresolved modeling or normative choices.

This distinction also makes \emph{Causal Cards} instruments of contestability rather than transparency alone. The card should allow another analyst, auditor, or stakeholder to identify where a conclusion could reasonably be challenged, for example because of an omitted mechanism, contested variable role, alternative causal specification, or disputed pathway classification. Contestability does not imply that every alternative receives equal weight; implausible specifications may be excluded on empirical or substantive grounds. What matters is that the basis for such exclusions remains visible.

Finally, a \emph{Causal Card} connects epistemic status to governance action. As argued above, uncertainty does not map mechanically onto a single response. A stable finding may support a stronger causal claim, while an assumption-sensitive or unresolved finding may warrant qualification, additional evidence, precaution, monitoring, or review depending on the stakes involved. Importantly, failure to establish unfairness under one causal specification should not itself be converted into a categorical claim that a system is fair.

Therefore, \emph{Causal Cards} function less as certificates than as records of accountable judgment. Their purpose is to preserve the distinction between what a causal analysis establishes and what an institution ultimately decides on its basis. The following section illustrates this function using alternative CD outputs from the UCI Adult dataset.

% \section{Worked Case: How Model Disagreement Produces Competing Accounts of Fairness}
\section{Worked Case: How Model Disagreement Structures Fairness Interpretation}
\label{sec:worked_case}

We illustrate the framework using the UCI Adult dataset \citep{kohavi1996scaling}. The purpose of the example is to demonstrate how different CD procedures applied to the same observational data can make different fairness-relevant explanations visible, absent, or unresolved.

The case therefore functions as an epistemic demonstration of how the causal representation selected for analysis can shape what an analyst is in a position to claim before any categorical fairness judgment is made.

\subsection{Case Setup}

The Adult dataset is commonly used to predict whether an individual's income exceeds \$50,000 based on attributes including education, occupation, work hours, marital status, employment characteristics, and demographic variables. We treat \texttt{Salary} as the outcome and highlight \texttt{Race}, \texttt{Sex}, \texttt{Age}, and \texttt{Marital Status} as fairness-relevant attributes.

The dataset is useful for this demonstration partly because of what it omits. Income may be shaped by mechanisms such as discrimination, labor-market segmentation, inherited wealth, institutional exclusion, and unequal access to educational or occupational opportunities that are not directly represented. Observed variables such as education, occupation, relationship status, workclass, and hours worked may therefore admit different causal and normative interpretations.

We apply three widely used CD procedures to the same observed feature set: Peter-Clark (PC) \citep{spirtes1991algorithm}, Fast Causal Inference (FCI) \citep{spirtes2001causation}, and Greedy Equivalence Search (GES) \citep{chickering2002optimal}. We examine directed paths from the fairness-relevant attributes to \texttt{Salary}. For FCI, only fully directed paths are interpreted as supporting a directed causal claim; partially oriented relationships are treated as unresolved.

\subsection{Alternative Causal Representations}

\begin{figure*}[ht]
    \centering
    \includegraphics[scale=0.090]{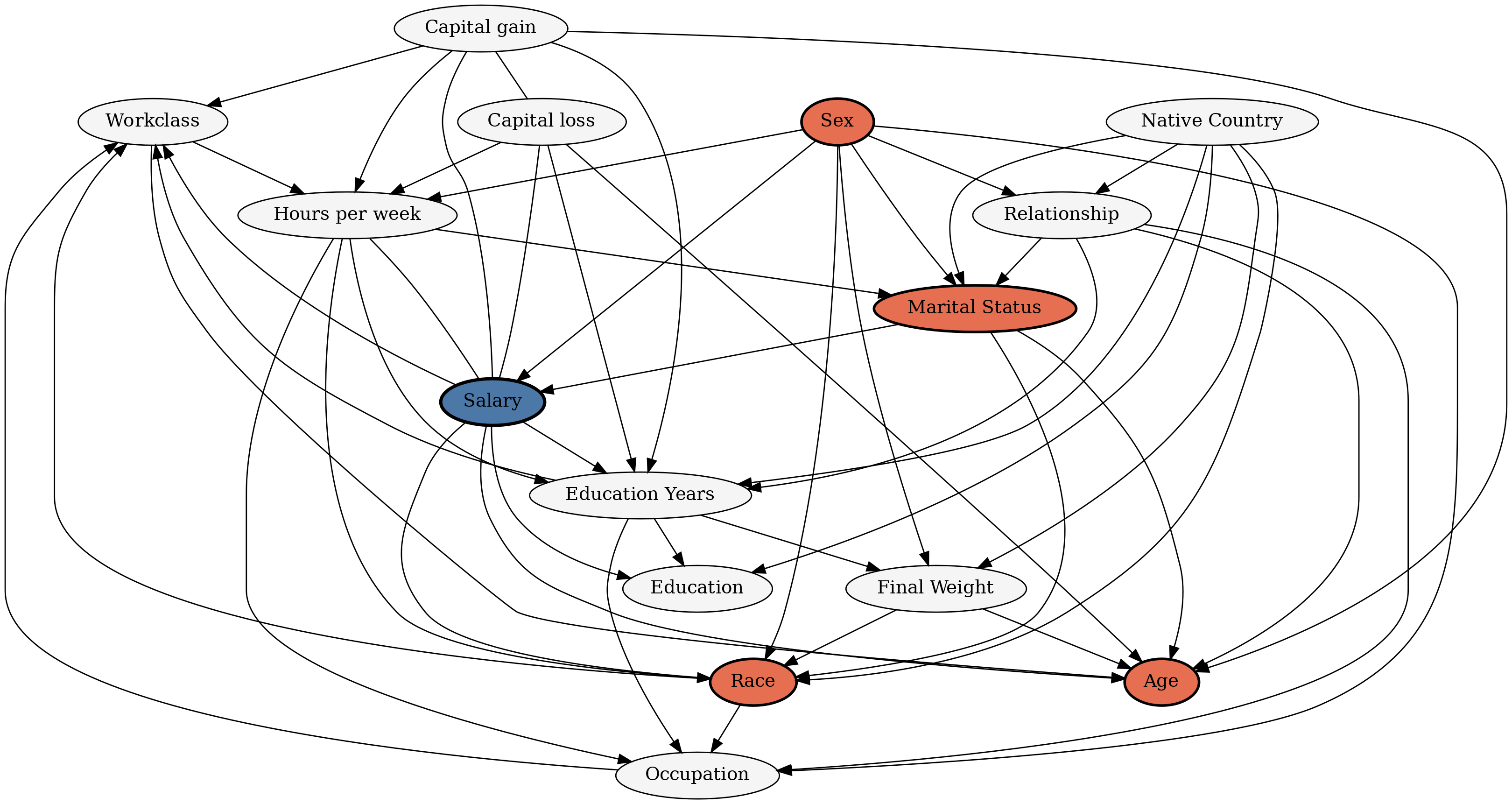}
    \caption{Causal structure learned from the UCI Adult dataset using PC. The fairness-relevant attributes (Race, Sex, Age, and Marital Status) and the outcome (Salary) are highlighted for emphasis.}
    \label{fig:adult_graphs_pc}
\end{figure*}

\begin{figure*}[ht]
    \centering
    \includegraphics[scale=0.37]{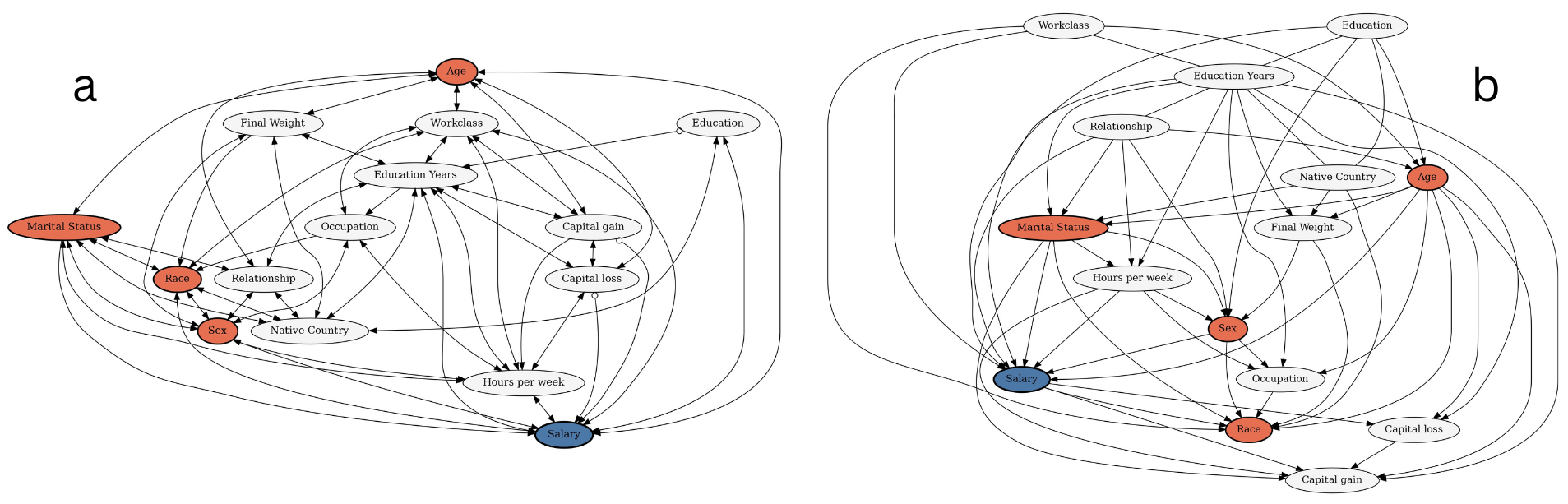}
    \caption{Causal structures learned from the UCI Adult dataset using (a) FCI and (b) GES. The fairness-relevant attributes (Race, Sex, Age, and Marital Status) and the outcome (Salary) are highlighted for emphasis. Partially oriented relationships in FCI are treated as unresolved rather than as confirmed directed pathways.}
    \label{fig:adult_graphs_fci_ges}
\end{figure*}

Figures~\ref{fig:adult_graphs_pc} and~\ref{fig:adult_graphs_fci_ges} show the structures returned by PC, FCI, and GES. The outputs differ substantially in graph density, edge orientation, and the pathways connecting fairness-relevant attributes to \texttt{Salary}.

Under PC, \texttt{Sex} has a direct edge to \texttt{Salary} and multiple indirect directed paths through variables including \texttt{Marital Status}, \texttt{Relationship}, \texttt{Workclass}, and \texttt{Hours per week}. This representation makes several sex-related mechanisms available for subsequent fairness interpretation, including questions about whether intermediate variables reflect legitimate determinants of income or prior inequality.

GES produces a substantially sparser representation. \texttt{Sex} retains a direct path to \texttt{Salary}, while \texttt{Race} has no directed path to the outcome. If interpreted in isolation, this structure would support a narrower set of race-related causal claims. Importantly, however, absence from the learned graph is not evidence that the corresponding social mechanism does not exist; it is an absence conditional on the observed variables and modeling procedure.

FCI leaves much of the fairness-relevant structure unresolved under our fully directed-path criterion. The only fully directed pathway among the highlighted attributes is from \texttt{Marital Status} to \texttt{Salary}. This does not establish that \texttt{Race}, \texttt{Sex}, or \texttt{Age} are causally irrelevant. Rather, the learned representation does not support a fully directed causal claim from those attributes to the outcome.

The comparison therefore produces three different epistemic possibilities: a fairness-relevant relationship may be \emph{visible} as an identified directed pathway, \emph{absent} from a particular learned representation, or \emph{unresolved} because the available structure does not support confident orientation. These distinctions matter because they structure the space of explanations available to a subsequent fairness assessment.

A statement such as ``no race-related causal pathway was identified under this specification,'' for example, is materially different from the stronger claim that race does not causally influence the outcome. The former describes the evidentiary output of a conditional analysis while the latter converts that output into a more general causal explanation. \emph{Causal humility} requires preserving this distinction.

\subsection{Applying a \emph{Causal Card}}

Table~\ref{tab:adult_causal_card} applies the \emph{Causal Card} framework to the example. Rather than selecting one graph as authoritative, the card records the alternative representations, consequential assumptions and omissions, and the model dependence of the resulting claim.

\begin{table*}[t]
\centering
\small
\caption{Illustrative Causal Card for the Adult income prediction example.}
\label{tab:adult_causal_card}
\begin{tabularx}{\textwidth}{p{0.24\textwidth} X}
\toprule
\textbf{Field} & \textbf{Illustrative entry} \\
\midrule

Decision context / claim &
Illustrative assessment of whether fairness-relevant attributes have identifiable causal pathways to an income-threshold outcome in the UCI Adult data. The example is not used to justify real-world deployment or certification. \\

\addlinespace

Outcome &
\texttt{Salary}, represented as an income-threshold indicator. \\

\addlinespace

Fairness-relevant attributes &
\texttt{Race}, \texttt{Sex}, \texttt{Age}, and \texttt{Marital Status}. \\

\addlinespace

Causal representations considered &
Structures learned from the same observed feature set using PC, GES, and FCI. The outputs are treated as alternative representations for comparison, not as equally valid ground-truth causal models. \\

\addlinespace

Key assumptions / analysis protocol &
Causal structure is inferred from observational data. Candidate pathways are traced from fairness-relevant attributes to \texttt{Salary}. Only fully directed paths are interpreted as directed causal pathways; partially oriented FCI relationships are treated as unresolved. \\

\addlinespace

Consequential omissions &
The observed data do not directly represent mechanisms including discrimination, labor-market segmentation, inherited wealth, institutional exclusion, historical disadvantage, or differential access to educational and occupational opportunities. \\

\addlinespace

Contested variable interpretations &
Variables including \texttt{Education Years}, \texttt{Occupation}, \texttt{Workclass}, \texttt{Relationship}, and \texttt{Marital Status} may have different causal or normative interpretations depending on the social account assumed. \\

\addlinespace

Model-dependent findings &
PC makes multiple directed sex-to-\texttt{Salary} pathways visible. GES retains a direct sex-to-\texttt{Salary} pathway but no directed race-to-\texttt{Salary} pathway. FCI leaves most highlighted relationships unresolved under the fully directed-path criterion. \\

\addlinespace

Epistemic status &
\emph{Assumption-sensitive / unresolved}. No protected-attribute-to-outcome pathway is consistently identified as a fully directed pathway across all three representations considered. \\

\addlinespace

Governance implication &
A categorical claim about the presence or absence of a discriminatory causal pathway should not be justified from any single discovery output without further evidence. Downstream use should preserve the dependence of the finding on model specification and relevant mechanisms omitted from the observed data. \\

\bottomrule
\end{tabularx}
\end{table*}

The \emph{Causal Card} changes the form of the resulting claim. A single-model analysis might conclude by reporting which protected attributes do or do not have identified pathways to \texttt{Salary}. The card instead records that these findings depend on the causal representation considered and that important social mechanisms remain outside the observed data. The appropriate output is therefore a characterization of what the available evidence can and cannot support, rather than a categorical fairness judgment.

\subsection{What the Worked Case Establishes}

The example supports a deliberately limited claim. It does not show that PC, GES, and FCI are equally credible accounts of the causal structure of income, nor that real institutions would necessarily reach different deployment decisions under each representation. Establishing those claims would require substantive domain validation and empirical study of institutional practice.

What the example does show is that causal representation structures fairness interpretation. Different discovery procedures applied to the same observational feature set can render fairness-relevant relationships visible, absent, or unresolved. When that model dependence is material to the fairness claim, it should remain part of the finding rather than disappear when the analysis is translated into a stronger institutional conclusion.

\section{Discussion}
\label{sec:discussion}
Taken together, the arguments in this paper suggest that the turn to causal fairness relocates rather than removes normative judgment from fairness analysis. Statistical fairness has long been criticized for embedding normative choices in decisions about which outcomes, error rates, or distributions should be equalized \citep{green2018myth,selbst2019fairness}. Causal approaches introduce additional sites of judgment: which mechanisms should be represented, how variables should be interpreted, which pathways count as permissible, and which causal alternatives are sufficiently credible to consider. Greater explanatory capacity therefore does not remove normative judgment from fairness evaluation; it relocates some of that judgment into the construction and interpretation of causal accounts.

This has implications for how expertise is organized around causal fairness assessments. If causal explanations can influence judgments about discrimination, responsibility, or acceptable intervention, determining whether a causal account is sufficiently credible cannot always be treated as a purely technical matter. Analysts may have expertise in identification or CD, while domain experts, auditors, affected stakeholders, and legal or policy actors may possess different forms of knowledge relevant to interpreting variables, omissions, or institutional processes. \emph{Causal humility} therefore raises an institutional question about how these forms of expertise should interact and who has standing to challenge a causal account. Contestability therefore requires institutional processes through which relevant challenges can be heard, evaluated, and incorporated, rather than assumption disclosure alone.

A related tension arises between epistemic qualification and institutional decisiveness. Audits and governance processes generally exist because organizations must eventually make decisions either to deploy, intervene, investigate, restrict, or accept a system. Yet the demand for an actionable conclusion can create pressure to resolve ambiguities that the underlying evidence has not resolved. The challenge is to make decisions under uncertainty without allowing the institutional need for decisiveness to retrospectively make the causal evidence appear more certain than it is. In this sense, \emph{causal humility} makes the basis and limits of judgment more explicit rather than attempting to eliminate judgment from governance.

Uncertainty itself may also become contested. Different actors may reasonably disagree about whether residual uncertainty is substantial enough to weaken a causal claim, warrant additional investigation, justify precaution, or permit continued use. Institutional incentives can further shape these judgments. Uncertainty may be minimized when certainty supports authorization and emphasized when ambiguity provides grounds to defer mitigation or responsibility. Therefore, governing causal uncertainty requires attention to how evidentiary standards are applied, by whom, and with what consequences. The central issue now becomes how the significance of uncertainty is determined in relation to the harms and decisions at stake.

These implications broaden the problem addressed by \emph{causal humility}. Responsible causal fairness depends on the quality of causal methods, but also on the institutional arrangements through which causal explanations are interpreted, challenged, and authorized. Seen in this way, causal fairness is also a data practice through which institutional actors transform observational evidence into authoritative accounts of inequality and its causes. This shifts part of the research agenda from improving causal identification alone toward examining the practices through which causal evidence acquires governance significance. Understanding those practices is important for ensuring that increasingly sophisticated causal analyses support more accountable fairness judgments rather than simply more authoritative ones.

\section{Limitations and Future Directions}
\label{sec:limitations}

This paper develops a conceptual and procedural framework for governing causal fairness claims under epistemic uncertainty. It does not resolve the underlying problems of causal identification. When relevant confounders are unobserved, causal directions remain ambiguous, or multiple causal structures are compatible with the available evidence, \emph{causal humility} cannot determine which causal account is correct. Its purpose is instead to constrain what can responsibly be claimed and how unresolved uncertainty is carried into subsequent governance decisions.

The worked case is similarly illustrative rather than validating. The PC, FCI, and GES outputs are used to demonstrate how different causal representations can make fairness-relevant pathways visible, absent, or unresolved; they are not established as equally plausible accounts of the true causal structure of income. Nor does the example apply a complete causal fairness metric or reproduce a production audit. Stronger empirical evaluation would require substantive domain knowledge, explicit justification of the admissible model set, and comparison against settings in which relevant causal relationships are better supported.

A further limitation concerns the specification of which alternative causal models or assumptions should count as sufficiently plausible for contestation. \emph{Causal humility} rejects both reliance on a single unexamined specification and indiscriminate comparison across arbitrary models, but it does not provide a universal rule for defining the relevant comparison set. In practice, this boundary will depend on empirical evidence, domain expertise, theoretical commitments, and the institutional purpose of the assessment. Developing more systematic procedures for identifying materially relevant alternatives is therefore an important direction for future work.

The framework also imposes practical demands. Documenting consequential assumptions, examining alternative specifications, and preserving epistemic qualifications across organizational processes require expertise, time, and institutional capacity. \emph{Causal Cards} are proposed as one mechanism for supporting these practices, but we do not empirically evaluate how analysts, auditors, decision-makers, or affected stakeholders would interpret or use them. Their usability, appropriate level of detail, and effectiveness in preventing the loss of material uncertainty remain open empirical questions.

Future work should evaluate \emph{causal humility} in real auditing and governance settings while developing stronger methodological links between technical uncertainty analysis and institutional decision-making. Empirical studies could examine how causal findings are translated across technical, compliance, legal, and managerial contexts; whether epistemic qualifications are preserved or lost; and whether structured artifacts such as \emph{Causal Cards} meaningfully influence institutional judgments. Participatory evaluation could further assess whether domain experts and affected stakeholders identify consequential omissions or interpretations that technical analyses overlook. In parallel, methods such as sensitivity analysis, partial identification, causal model averaging, and explicit representation of graph uncertainty could provide more systematic ways to characterize the stability of fairness-relevant claims. Connecting these technical approaches to governance processes that account for harm severity, reversibility, and institutional responsibility would help assess both the practical utility of \emph{Causal Cards} and the broader feasibility of \emph{causal humility} as an institutional practice.

\section{Conclusion}

Causal approaches to algorithmic fairness promise richer explanations of why disparities arise, but those explanations depend on causal representations that may remain partial, assumption-sensitive, or contested. This paper argued that the resulting problem is not only one of causal uncertainty, but of \emph{causal authority}: uncertain causal accounts can acquire institutional force when they are used to interpret disparities, justify interventions, or support judgments about AI systems.

We introduced \emph{causal humility} as a governance principle for keeping the authority of fairness claims proportionate to their epistemic support. By making consequential assumptions visible, preserving contestability and indeterminacy, and linking epistemic support to institutional action, \emph{causal humility} seeks to retain the value of causal reasoning without turning conditional evidence into categorical judgment. \emph{Causal Cards} offer one mechanism for carrying these qualifications into governance processes. More broadly, responsible causal fairness requires attention to what causal models reveal and to what they omit, how their claims acquire authority, and how uncertainty is preserved as those claims move into institutional decision-making.

\bibliographystyle{SageH}
\bibliography{sagebib.bib}

\end{document}